\documentclass[twoside,twocolumn,english,prb,showpacs]{revtex4}
\usepackage[T1]{fontenc}
\usepackage[latin1]{inputenc}
\usepackage{graphicx}
\usepackage{amssymb}

\makeatletter

\usepackage{babel}
\makeatother
\begin{document}

\title{Quantum rotor description of the Mott-insulator transition in the
Bose-Hubbard model}

\begin{abstract}
We present the novel approach to the Bose-Hubbard model using the
$\mathrm{U}\left(1\right)$ quantum rotor description. The effective
action formalism allows us to formulate a problem in the phase only
action and obtain an analytical formulas for the critical lines. We
show that the nontrivial $\mathrm{U}\left(1\right)$ phase field configurations
have an impact on the phase diagrams. The topological character of
the quantum field is governed by terms of the integer charges - winding
numbers. The comparison presented results to recently obtained quantum
Monte Carlo numerical calculations suggests that the competition between
quantum effects in strongly interacting boson systems is correctly
captured by our model. 
\end{abstract}

\author{T. P. Polak and T. K. Kope\'c}

\address{Institute for Low Temperatures and Structure Research, Polish Academy
of Sciences, POB 1410, 50-950 Wroclaw 2, Poland}


\maketitle

\section{Introduction}

The physics of the Bose-Hubbard (BH) model was the subject of intensive
study for a number of years.\cite{fisher,sachdev} More recently it
has been realized that the Bose-Hubbard model can also be applied
to bosons trapped in so-called optical lattices,\cite{jaksch,greiner}
and coarse graining\cite{kampf}, strong-coupling expansion,\cite{freericks,elstner}
mean-field theories\cite{sheshadri} have been successfully applied
to these systems in one-,\cite{clark,damski} two-\cite{niemeyer}
and three-dimensional lattices.\cite{sewer} Another essentially equivalent
formulation is based on the Gutzwiller wavefunction.\cite{krauth,rokshar}
Also extension, based on a systematic strong-coupling approach of
the Bose-Hubbard model beyond mean-field has been tried\cite{sengupta}
and experimentally confirmed.\cite{spielman} The progress comes from
better computer resources and more efficient algorithm allows to use
the Quantum Monte Carlo (QMC) method for studies of the BH systems.\cite{prokofev,alet,caproso} 

An optical lattices offer remarkably clean access to a particular
Hamiltonian and thereby serve as a model system for testing fundamental
theoretical concepts and providing exemplar of quantum many-body effects.\cite{bloch}.
It is well known that the ground state of a system of repulsively
interacting bosons in a periodic potential can be either in a superfluid
state (SF) or in a Mott-insulting state (MI), characterized by integer
boson densities. Because the phase of the order parameter and the
particle number as conjugate variables are subject to the uncertainty
principle $\Delta\phi\Delta n\sim\hbar$\cite{elion} and the bosons
can either be in the eigenstate of particle number or phase. The eigenstate
of phase is a superfluid and that of particle number is a localized
Mott insulator. Therefore the quantum MI-SF phase transition takes
place as the particle density is shifted thus facilitating emergence
of the superfluid from the Mott insulating state.

The aim of this paper is to extend the mean-field approach for the
Bose-Hubbard model in a way to include particle number fluctuations
effects and make the qualitative phase diagrams in two and three dimensions
more quantitative. Our method also improve the strong-coupling expansion
that works well only for sufficiently large insulating gap.\cite{caproso}
The key point of presented approach is to consider the representation
of strongly interacting bosons as particles with attached \char`\"{}flux
tubes\char`\"{}. In a consequence a boson is the composite object.
This introduces a conjugate $\mathrm{U}\left(1\right)$ phase variable,
which acquires dynamic significance from the boson-boson interaction.
To facilitate this task we employ the functional integral formulation
of the theory that enables us to perform the functional integration
over fields defined on different topologically equivalent classes
of the $\mathrm{U}\left(1\right)$ group, i.e., with different winding
numbers. An inclusion of the winding numbers is unavoidable in order
to obtain a proper phase diagram. A similar method that is based on
quantum rotor formulation was recently employed by one of us in the
fermionic Hubbard model.\cite{kopec} The nice feature of our approach
is that all the expressions and handling are analytic. Finally, we
compare our results for systems at zero temperature with the outcome
of the numerical simulations and found a very good agreement for the
quantitative results regarding the behavior as we go from the superfluid
phase to the Mott insulating phase. In the framework of the introduced
theory we are able to calculate the phase diagrams with high accuracy
along whole critical line that separates Mott insulator - superfluid
phases. Moreover, our approach gives a clear insight into the Bose-Hubbard
model from a quantum field theory and emphasizes the impact of the
topology of the phase variable on phase transitions. We show that
the Coulomb interaction (as a main energy scale) is governed the phase
transitions in the Bose-Hubbard model. 

The outline of the paper is as follows. In Sec. II we introduce the
model Hamiltonian and in Sec. III we derive an effective $\mathrm{U}\left(1\right)$
action in the quantum rotor representation. The aim of Sec. IV is
the presentation of the resulting phase diagrams for two- and three-dimensional
Bose-Hubbard systems. Finally, Section V summarizes our results and
sets the outlook.

\section{Model Hamiltonian}

We investigate the generic model for the Mott-insulator transition
the Bose-Hubbard model\begin{equation}
\mathcal{H}=\frac{U}{2}\sum_{i}n_{i}\left(n_{i}-1\right)-\sum_{\left\langle i,j\right\rangle }t_{ij}a_{i}^{\dagger}a_{j}-\mu\sum_{i}n_{i},\label{hamiltonian1}\end{equation}
where $a_{i}^{\dagger}$ and $a_{j}$ stands for the bosonic creation
and annihilation operators that obey the canonical commutation relations
$\left[a_{i},a_{j}^{\dagger}\right]=\delta_{ij}$, $n_{i}=a_{i}^{\dagger}a_{i}$
is the boson number operator on the site $i$, $U>0$ is the on-site
repulsion and $\mu$ is the chemical potential which controls the
number of bosons. Here, $\left\langle i,j\right\rangle $ identifies
summation over the nearest-neighbor sites. Furthermore, $t_{ij}$
is the hopping matrix element with the dispersion for the bipartite
lattice \begin{equation}
t\left(\mathbf{k}\right)=2t\sum_{l=1}^{d}\cos k_{l}.\label{dispersion}\end{equation}
in $d$ dimensions. In this paper we investigate the phase transitions
in simple cubic and square lattice. For our purpose we rewrite Eq.
(\ref{hamiltonian1}) to more suitable form\begin{eqnarray}
\mathcal{H} & = & \frac{U}{2}\sum_{i}n_{i}^{2}-\sum_{\left\langle i,j\right\rangle }t_{ij}a_{i}^{\dagger}a_{j}-\bar{\mu}\sum_{i}n_{i},\label{hamiltonian2}\end{eqnarray}
where $\bar{\mu}/U=\mu/U+1/2$ is the shifted reduced chemical potential.

\section{Method}

\subsection{Decoupling of the Coulomb interaction}

We will adopt the method of the quantum rotor model, developed by
one of us,\cite{kopec} to the BH Hamiltonian. The partition function
of the system could be written in the form\begin{equation}
\mathcal{Z}=\int\left[\mathcal{D}\bar{a}\mathcal{D}a\right]e^{-\mathcal{S}\left[\bar{a},a\right]}\end{equation}
and the bosonic path-integral is taken over the complex fields $a_{i}\left(\tau\right)$
with the action $\mathcal{S}$ given by\begin{equation}
\mathcal{S}=\mathcal{S}_{B}\left[\bar{a},a\right]+\int_{0}^{\beta}d\tau\mathcal{H\left(\tau\right)},\end{equation}
where\begin{equation}
\mathcal{S}_{B}\left[\bar{a},a\right]=\sum_{i}\int_{0}^{\beta}d\tau\bar{a}_{i}\left(\tau\right)\frac{\partial}{\partial\tau}a_{i}\left(\tau\right).\end{equation}
Unfortunately Hamiltonian is not quadratic in $a_{i}$ and we have
to decouple first - the Coulomb term in Eq. (\ref{hamiltonian2})
by a Gaussian integration over the auxiliary fields $V_{i}\left(\tau\right)$.
The transformed partition function becomes\begin{equation}
\mathcal{Z}=\int\left[\mathcal{D}\bar{a}\mathcal{D}a\right]e^{-\mathcal{S}_{1}\left[\bar{a},a\right]}\int\left[\frac{dV}{2\pi}\right]e^{-\mathcal{S}_{2}\left[n,V\right]},\end{equation}
where\begin{eqnarray}
\mathcal{S}_{1}\left[\bar{a},a\right] & = & \int_{0}^{\beta}d\tau\left[\sum_{i}\bar{a}_{i}\left(\tau\right)\frac{\partial}{\partial\tau}a_{i}\left(\tau\right)\right.\nonumber \\
 &  & \left.-\sum_{\left\langle i,j\right\rangle }t_{ij}\bar{a}_{i}\left(\tau\right)a_{j}\left(\tau\right)\right],\end{eqnarray}
and\begin{equation}
\mathcal{S}_{2}\left[n,V\right]=\sum_{i}\int_{0}^{\beta}d\tau\left\{ \frac{1}{2U}V_{i}^{2}\left(\tau\right)-\left[iV_{i}\left(\tau\right)-\bar{\mu}\right]n_{i}\left(\tau\right)\right\} .\end{equation}
After changing variables $V_{i}\left(\tau\right)=V_{i}^{T}\left(\tau\right)+\frac{1}{i}\bar{\mu}$
the second part of the action takes form\begin{eqnarray}
\mathcal{S}_{2}\left[n,V^{T}\right] & = & \sum_{i}\int_{0}^{\beta}d\tau\left\{ \frac{1}{2U}\left[V_{i}^{T}\left(\tau\right)\right]^{2}+\frac{\bar{\mu}}{iU}V_{i}^{T}\left(\tau\right)\right.\nonumber \\
 &  & \left.-\frac{\bar{\mu}^{2}}{2U}-iV_{i}^{T}\left(\tau\right)n_{i}\left(\tau\right)\right\} .\end{eqnarray}
The field $V_{i}^{T}\left(\tau\right)$ could be represent as a sum
of the static $V_{i}^{S}\left(\tau\right)$ and periodic function
$V_{i}^{P}\left(\tau\right)$:\begin{eqnarray}
V_{i}^{T}\left(\tau\right) & = & V_{i}^{S}\left(\tau\right)+V_{i}^{P}\left(\tau\right),\nonumber \\
V_{i}^{P}\left(\tau\right) & = & \frac{1}{\beta}\sum_{m=1}^{+\infty}\left[V_{i}^{P}\left(\omega\right)e^{i\omega_{m}\tau}+c.c.\right],\nonumber \\
V_{i}^{S} & = & \frac{1}{\beta}V_{i}^{T}\left(\omega_{m=0}\right).\label{electrochemical potential}\end{eqnarray}
where $\omega_{m}=2\pi m/\beta$ $\left(m=\pm1,\pm2..\right)$ are
the Bose-Matsubara frequencies. Furthermore, we introduce the scalar
potential field which couples to the local particle number through
the Josephson-like relation\begin{equation}
\dot{\phi}_{i}\left(\tau\right)=V_{i}^{P}\left(\tau\right),\label{josephson relation}\end{equation}
where the phase field satisfies the periodicity condition $\phi_{i}\left(\beta\right)=\phi_{i}\left(0\right)$
as a consequence of the periodic properties of the $V_{i}^{P}\left(\tau\right)$
field. We can eliminate the periodic parts of the fluctuating electrochemical
potential $V_{i}^{P}\left(\tau\right)$ from the action replacing
them by the phase field $\dot{\phi}_{i}\left(\tau\right)$:\begin{eqnarray}
\mathcal{Z} & = & \int\left[\mathcal{D}\bar{a}\mathcal{D}a\right]e^{-\mathcal{S}_{1}\left[\bar{a},a\right]}\nonumber \\
 & \times & \int\left[\frac{dV^{S}}{2\pi}\right]e^{-\mathcal{S}_{2}\left[n,V^{S}\right]}\int\left[\mathcal{D}\phi\right]e^{-\mathcal{S}_{3}\left[n,\dot{\phi}\right]},\label{partition function}\end{eqnarray}
where\begin{eqnarray}
\mathcal{S}_{2}\left[n,V^{S}\right] & = & \beta\sum_{i}\left[\frac{1}{2U}\left(V_{i}^{S}\right)^{2}+\frac{\bar{\mu}}{iU}V_{i}^{S}\right.\nonumber \\
 &  & \left.-\frac{iV_{i}^{S}}{\beta}\int_{0}^{\beta}d\tau n_{i}\left(\tau\right)-\frac{\bar{\mu}^{2}}{2U}\right],\\
\mathcal{S}_{3}\left[n,\dot{\phi}\right] & = & \sum_{i}\int_{0}^{\beta}d\tau\left[\frac{1}{2U}\dot{\phi_{i}^{2}}\left(\tau\right)+\frac{\bar{\mu}}{iU}\dot{\phi_{i}}\left(\tau\right)\right.\nonumber \\
 &  & \left.-i\dot{\phi_{i}}\left(\tau\right)n_{i}\left(\tau\right)\right].\end{eqnarray}
The factor with $-i\int_{0}^{\beta}d\tau\dot{\phi_{i}}\left(\tau\right)n_{i}\left(\tau\right)$
can be removed from the last equation by performing the local gauge
transformation to the new bosonic variables as we show in the next
subsection.

\subsection{Gauge transformation}

We perform the local gauge transformation to the new bosonic variables\begin{equation}
\left[\begin{array}{c}
a_{i}\left(\tau\right)\\
\bar{a}_{i}\left(\tau\right)\end{array}\right]=\left[\begin{array}{cc}
e^{i\phi_{i}\left(\tau\right)} & 0\\
0 & e^{-i\phi_{i}\left(\tau\right)}\end{array}\right]\left[\begin{array}{c}
b_{i}\left(\tau\right)\\
\bar{b}_{i}\left(\tau\right)\end{array}\right].\label{gauge}\end{equation}
The $\mathrm{U}\left(1\right)$ group governing the phase field is
compact, i.e. $\phi\left(\tau\right)$ has the topology of a circle
$S_{1}$, so that instanton effects can arise due to non-homotopic
mappings of the configuration space onto the gauge group $\mathrm{U}\left(1\right)$.
Therefore, we concentrate on closed paths in the imaginary time $\left(0,1/k_{\mathrm{B}}T\right)$
which fall into distinct, disconnected (homotopy) classes labelled
by the integer winding numbers $n_{i}$.\cite{rivers} The chief merit
of the transformation in Eq. (\ref{gauge}) is that we have managed
to cast the strongly correlated bosonic problem into a system of weakly
interacting bosons, submerged in the bath of strongly fluctuating
$\mathrm{U}\left(1\right)$ gauge potentials (on the high energy scale
set by $U$). Now the action contains three parts: \begin{eqnarray}
\mathcal{S}_{1}\left[\bar{b},b,\phi\right] & = & \int_{0}^{\beta}d\tau\left\{ \sum_{i}\bar{b}_{i}\left(\tau\right)\frac{\partial}{\partial\tau}b_{i}\left(\tau\right)\right.\nonumber \\
 & - & \left.\sum_{\left\langle i,j\right\rangle }t_{ij}\bar{b}_{i}\left(\tau\right)b_{j}\left(\tau\right)e^{-i\phi_{ij}\left(\tau\right)}\right\} ,\\
\mathcal{S}_{2}\left[V^{S}\right] & = & \beta\sum_{i}\left[\frac{1}{2U}\left(V_{i}^{S}\right)^{2}+\frac{\bar{\mu}}{iU}V_{i}^{S}\right.\nonumber \\
 & - & \left.\frac{iV_{i}^{S}}{\beta}\int_{0}^{\beta}d\tau\bar{b}_{i}\left(\tau\right)b_{i}\left(\tau\right)-\frac{\bar{\mu}^{2}}{2U}\right],\\
\mathcal{S}_{3}\left[\dot{\phi}\right] & = & \sum_{i}\int_{0}^{\beta}d\tau\left\{ \frac{1}{2U}\dot{\phi_{i}^{2}}\left(\tau\right)+\frac{1}{i}\frac{\bar{\mu}}{U}\dot{\phi_{i}}\right\} ,\end{eqnarray}
where $\phi_{ij}\left(\tau\right)=\phi_{i}\left(\tau\right)-\phi_{j}\left(\tau\right)$
and still we have terms with $V_{i}^{S}$ that will be calculated
in the next subsection. Furthermore, the path-integral includes a
summation over winding numbers\begin{equation}
\int\left[\mathcal{D}\phi\right]...\equiv\sum_{\left\{ n_{i}\right\} }\int_{0}^{2\pi}\prod_{i}d\phi_{i}\left(0\right)\int_{_{\phi_{i}\left(0\right)}}^{\phi\left(\tau\right)_{i}+2\pi n_{i}}\prod_{i}d\phi_{i}\left(\tau\right)...\end{equation}
and should be performed taking phase configurations that satisfy boundary
condition $\phi_{i}\left(\beta\right)-\phi_{i}\left(0\right)=2\pi n_{i}$
where $n_{i}$ is integer.

\subsection{Saddle point equation}

The expectation value of the static part of the fluctuating electrochemical
potential\begin{equation}
\left\langle V^{S}\right\rangle =\frac{\int\left[\mathcal{D}V^{S}\right]V^{S}e^{-\mathcal{S}_{2}\left[V^{S}\right]}}{\int\left[\mathcal{D}V^{S}\right]e^{-\mathcal{S}_{2}\left[V^{S}\right]}}\label{integral}\end{equation}
 introduced in Eq. (\ref{electrochemical potential}) we calculate
using the saddle point approximation and for $U>0$ obtain:\begin{equation}
V^{S}=i\left[\bar{\mu}+U\left\langle \bar{b}_{i}\left(\tau\right)b_{i}\left(\tau\right)\right\rangle \right].\label{static part}\end{equation}
Now making substitution in the second part of the action $\mathcal{S}_{2}\left[V^{S}\right]$
for the $V_{i}^{S}$ an unique global value obtained from Eq. (\ref{static part})
we get finally\begin{eqnarray}
\mathcal{S}_{2}\left[\bar{b},b\right] & = & \beta\sum_{i}\frac{U}{2}\left\langle \bar{b}_{i}\left(\tau\right)b_{i}\left(\tau\right)\right\rangle ^{2}\nonumber \\
 & + & \bar{\mu}\sum_{i}\int_{0}^{\beta}d\tau\bar{b}_{i}\left(\tau\right)b_{i}\left(\tau\right).\label{chemical potential action}\end{eqnarray}
The effective action is now quadratic in the bosonic variables and
can be integrated out without any difficulty remembering that the
first term in Eq. (\ref{chemical potential action}) is simply a number.
Therefore, the applied steepest descent method used to approximate
integral Eq. (\ref{integral}) allowed us to remove an after effects
of the auxiliary fields $V_{i}\left(\tau\right)$ introduced in order
to decouple the non-quadratic terms (in $a_{i}$ ) in Hamiltonian
Eq. (\ref{hamiltonian2}).

\subsection{The partition function expressed in the phase fields variables}

The partition function can be expressed in form of the effective propagator
$\hat{G}$:\begin{equation}
\mathcal{Z}=\int\left[\mathcal{D}\phi\right]e^{\left[-\sum_{i}\int_{0}^{\beta}d\tau\left(\frac{1}{2U}\dot{\phi_{i}^{2}}\left(\tau\right)+\frac{1}{i}\frac{\bar{\mu}}{U}\dot{\phi_{i}}\right)+\mathrm{Tr}\ln\hat{G}^{-1}\right]},\label{partition function propagator}\end{equation}
where $\exp\left(-\mathrm{Tr}\ln\hat{G}^{-1}\right)\equiv\det\hat{G}$
and determinant takes form\begin{eqnarray}
\det\hat{G} & = & \int\left[\mathcal{D}\bar{b}\mathcal{D}b\right]\exp\left\{ -\sum_{\left\langle i,j\right\rangle }\int_{0}^{\beta}d\tau\right.\nonumber \\
 & \times & \left.\bar{b}_{i}\left[\left(\frac{\partial}{\partial\tau}+\bar{\mu}\right)\delta_{ij}-t_{ij}e^{-i\phi_{ij}\left(\tau\right)}\right]b_{i}\right\} .\end{eqnarray}
We parametrize the boson fields \begin{equation}
b_{i}\left(\tau\right)=b_{0}+b_{i}^{'}\left(\tau\right)\label{parametrization}\end{equation}
 and restrict our calculations to the phase fluctuations dropping
the amplitude dependence. In result the inverse of the propagator
becomes\begin{equation}
\hat{G}^{-1}=\hat{G}_{0}^{-1}-T=\hat{G}_{0}^{-1}\left(1-T\hat{G}_{0}\right).\end{equation}
The explicit value $b_{0}$ can be obtained from minimalization of
the Hamiltonian $\partial\mathcal{H}\left(b_{0}\right)/\partial b_{0}=0$
where we introduced the parametrization Eq. (\ref{parametrization}).
Therefore, we write\begin{eqnarray}
\hat{G}_{0} & = & b_{0}^{2}\equiv\frac{\sum_{\left\langle i,j\right\rangle }t_{ij}+\bar{\mu}}{U},\label{zero mode}\\
T & = & t_{ij}e^{-i\phi_{ij}\left(\tau\right)}.\label{hopping}\end{eqnarray}
Kampf and Zimanyi\cite{kampf} considered similar parametrization
in the path-integral formulation of the coarse-graining procedure
to the BH model. However, to obtain a critical line, authors used
a mean-field approach that is not expected to be reliable at $T=0$
and be capable to handle spatial and quantum fluctuation effects properly,
especially in two dimensions. Moreover, as we will see in the next
sections, our results strongly depend on the dimension of the system
giving qualitative changing of the phase diagrams. 

Expanding the trace of the logarithm we have \begin{eqnarray}
\mathrm{Tr}\ln\hat{G}^{-1} & = & -\mathrm{Tr}\left(\ln\hat{G}_{0}\right)-\mathrm{Tr}\left(T\hat{G}_{0}\right)\nonumber \\
 &  & -\frac{1}{2}\mathrm{Tr}\left[\left(T\hat{G}_{0}\right)^{2}\right]+...\end{eqnarray}
with $\hat{G}_{0}$ and $T$ given by Eq. (\ref{zero mode}) and (\ref{hopping}).
Trace over first term of the expansion gives us constant contribution
to the action. From the trace over second part \begin{equation}
\mathrm{Tr}\left(T\hat{G}_{0}\right)=\sum_{\left\langle i,j\right\rangle }J_{ij}\int_{0}^{\beta}d\tau\cos\left[\phi_{i}\left(\tau\right)-\phi_{j}\left(\tau\right)\right]\end{equation}
we get the explicit form of the coefficient:\begin{equation}
J_{ij}=b_{0}^{2}t_{ij}=\frac{\sum_{\left\langle i,j\right\rangle }t_{ij}+\bar{\mu}}{U}t_{ij}.\end{equation}
 Finally a partition function Eq. (\ref{partition function propagator})
becomes\begin{eqnarray}
\mathcal{Z} & = & \int\left[\mathcal{D}\phi\right]e^{-\mathcal{S}_{\mathrm{ph}}\left[\phi\right]}\end{eqnarray}
with an effective action expressed only in the phase fields variable\begin{eqnarray}
\mathcal{S}_{\mathrm{ph}}\left[\phi\right] & = & \int_{0}^{\beta}d\tau\left\{ \sum_{i}\left[\frac{1}{2U}\dot{\phi_{i}^{2}}\left(\tau\right)+\frac{1}{i}\frac{\bar{\mu}}{U}\dot{\phi_{i}}\left(\tau\right)\right]\right.\nonumber \\
 &  & \left.-J\sum_{i,j}e^{\phi_{i}\left(\tau\right)}\mathcal{I}_{ij}e^{\phi_{j}\left(\tau\right)}\right\} ,\label{action only phase}\end{eqnarray}
where $\mathcal{I}_{ij}=1$ if $i,j$ are the nearest neighbors and
equals zero otherwise. 

To proceed we replace the phase degrees of freedom by the complex
field $\psi_{i}$ which satisfies the quantum periodic boundary condition
$\psi_{i}\left(\beta\right)=\psi_{i}\left(0\right)$. This can be
conveniently done using the Fadeev-Popov method with Dirac delta functional
representation in a way used by Kope\'c\cite{kopec1}:\begin{eqnarray}
1 & = & \int\left[\mathcal{D}\psi_{i}\mathcal{D}\psi_{i}^{*}\right]\delta\left(\sum_{i}\left|\psi\left(\tau\right)\right|^{2}-N\right)\nonumber \\
 & \times & \delta\left(\psi_{i}-e^{i\phi_{i}\left(\tau\right)}\right)\delta\left(\psi_{i}^{*}-e^{-i\phi_{i}\left(\tau\right)}\right).\label{popov}\end{eqnarray}
The main idea of this approach is to attempt to generate an effective
partition function from the original one with cosine interaction,
which incorporates the constrained nature of the original variables.
Thus we take $\psi_{i}$ as continuous variable but constrained (on
the average) to have the unit length:\begin{eqnarray}
\delta\left(\sum_{i}\left|\psi_{i}\left(\tau\right)\right|^{2}-N\right) & = & \frac{1}{2\pi i}\int_{-i\infty}^{+i\infty}d\lambda\nonumber \\
 & \times & e^{\int_{0}^{\beta}d\tau\lambda\left(\sum_{i}\left|\psi_{i}\left(\tau\right)\right|^{2}-N\right)}.\label{spherical constraint}\end{eqnarray}
In Eq. (\ref{spherical constraint}) we introduced the Lagrange multiplier
$\lambda$ which adds the quadratic terms (in the $\psi_{i}$ fields)
to the action Eq. (\ref{action only phase}). Using such description
is justified by the definition of the order parameter\begin{equation}
\Psi_{B}=\left\langle e^{i\phi_{i}\left(\tau\right)}\right\rangle \end{equation}
 which non-vanishing value signals a macroscopic quantum phase coherence
(in our case we identify it as the SF state). The partition function
can be written in form\begin{eqnarray}
\mathcal{Z} & = & \frac{1}{2\pi i}\int_{-i\infty}^{+i\infty}d\lambda\int\left[\mathcal{D}\psi_{i}\mathcal{D}\psi_{i}^{*}\right]e^{-\mathcal{S}_{\mathrm{eff}}},\end{eqnarray}
 where effective action $\mathcal{S}_{\mathrm{eff}}$ is given by:\begin{eqnarray}
\mathcal{S}_{\mathrm{eff}} & = & \sum_{i,j}\int_{0}^{\beta}d\tau d\tau^{'}\left[\left(J\mathcal{I}_{ij}+\lambda\delta_{ij}\right)\delta\left(\tau-\tau'\right)\right.\nonumber \\
 & - & \left.\mathcal{\gamma}_{ij}\left(\tau,\tau'\right)\right]\psi_{i}\psi_{j}^{*}-N\lambda\delta\left(\tau-\tau'\right).\end{eqnarray}
 Here\begin{eqnarray}
\gamma_{ij}\left(\tau,\tau'\right) & = & \left\langle e^{-i\left[\phi_{i}\left(\tau\right)-\phi_{j}\left(\tau'\right)\right]}\right\rangle \end{eqnarray}
is the two-point phase correlator associated with the order parameter
field. Summarizing this part, we formulated a problem introducing
an appropriate constrained complex order parameter field. In the next
section we show that the presence of the nontrivial topology possessed
by the phase variable will contribute to propagator.

\subsection{Topological contribution in the correlation function}

The existence of the topological features of the charge states affects
the correlation function. Because the values of the phases $\phi_{i}$
which differ by $2\pi$ are equivalent thus we decompose the phase
field in terms of a periodic field and term linear in $\tau$:\begin{equation}
\phi_{i}\left(\tau\right)=\varphi_{i}\left(\tau\right)+\frac{2\pi}{\beta}n_{i}\tau\end{equation}
with $\phi_{i}\left(\beta\right)=\phi_{i}\left(0\right).$ As a result
the phase correlator factorizes as the product of a topological term
$\gamma_{i}^{T}\left(\tau,\tau'\right)$ depending on the integers
$n_{i}$ and non-topological one $\gamma_{ij}^{N}\left(\tau,\tau'\right)$:\begin{equation}
\gamma_{ij}\left(\tau,\tau'\right)=\gamma_{i}^{T}\left(\tau,\tau'\right)\gamma_{ij}^{N}\left(\tau,\tau'\right)\end{equation}
where\begin{equation}
\gamma_{i}^{T}\left(\tau,\tau'\right)=\frac{\sum_{\left[n_{i}\right]}e^{-i\frac{2\pi}{\beta}\left(\tau-\tau'\right)n_{i}}e^{-\frac{2\pi}{\beta}\sum_{i}\left[\frac{\pi}{U}n_{i}^{2}+\frac{\beta}{i}\frac{\bar{\mu}}{U}n_{i}\right]}}{\sum_{\left[n_{i}\right]}e^{-\frac{2\pi}{\beta}\sum_{i}\left[\frac{\pi}{U}n_{i}^{2}+\frac{\beta}{i}\frac{\bar{\mu}}{U}n_{i}\right]}}\end{equation}
and\begin{equation}
\gamma_{ij}^{N}\left(\tau,\tau'\right)=\frac{\int\left[\mathcal{D}\varphi_{i}\right]e^{-i\left[\varphi_{i}\left(\tau\right)-\varphi_{j}\left(\tau^{'}\right)\right]}e^{-\sum_{i}\frac{1}{2U}\int_{0}^{\beta}d\tau\dot{\varphi_{i}^{2}}\left(\tau\right)}}{\int\left[\mathcal{D}\varphi_{i}\right]e^{-\sum_{i}\frac{1}{2U}\int_{0}^{\beta}d\tau\dot{\varphi_{i}^{2}}\left(\tau\right)}}.\end{equation}
Performing the Poisson re-summation formula\begin{equation}
\left|\sqrt{\det\mathbf{G}}\right|\sum_{\left[n_{i}\right]}e^{-\pi\left(n-a\right)_{i}\mathbf{G}_{ij}\left(n-a\right)_{j}}=\sum_{\left[n_{i}\right]}e^{-\pi m_{i}\left(\mathbf{G}^{-1}\right)_{ij}m_{j}}\end{equation}
 in $\gamma_{i}^{T}\left(\tau,\tau'\right)$ and the functional integration
over the phase variables in $\gamma_{ij}^{N}\left(\tau,\tau'\right)$
the final form of the correlator\begin{eqnarray}
\gamma_{ij}\left(\tau,\tau'\right) & = & \delta_{ij}e^{\frac{U}{2}\left|\tau-\tau^{'}\right|}\nonumber \\
 & \times & \frac{\sum_{\left[n_{i}\right]}e^{-\frac{U\beta}{2}\left(n_{i}+\frac{\bar{\mu}}{U}\right)^{2}}e^{-U\left(n_{i}+\frac{\bar{\mu}}{U}\right)\left(\tau-\tau'\right)}}{\sum_{\left[n_{i}\right]}e^{-\frac{U\beta}{2}\left(n_{i}+\frac{\bar{\mu}}{U}\right)^{2}}}\end{eqnarray}
 after Fourier transform can be written as:\begin{equation}
\gamma\left(\omega_{m}\right)=\frac{1}{\mathcal{Z}_{0}}\frac{4}{U}\sum_{\left[n_{i}\right]}\frac{e^{-\frac{U\beta}{2}\sum_{i}\left(n_{i}+\frac{\bar{\mu}}{U}\right)^{2}}}{1-4\left[\sum_{i}n_{i}+\frac{\bar{\mu}}{U}-i\frac{\omega_{m}}{U}\right]^{2}},\label{correlator}\end{equation}
where\begin{equation}
\mathcal{Z}_{0}=\frac{\delta_{ij}}{\sum_{\left[n_{i}\right]}e^{-\frac{U\beta}{2}\sum_{i}\left(n_{i}+\frac{\bar{\mu}}{U}\right)^{2}}}\end{equation}
is the partition function for the set of quantum rotors. The form
of Eq. (\ref{correlator}) assures the periodicity in the imaginary
time. We want to point out that another important property possessed
by correlator Eq. (\ref{correlator}) comes out. Namely, the propagator
is periodic with respect to $\mu/U+1/2$ which emphasizes the special
role of its integer values.

\section{Mott insulator - superfluid phase transition}

\begin{figure}
\includegraphics[%
  scale=0.43]{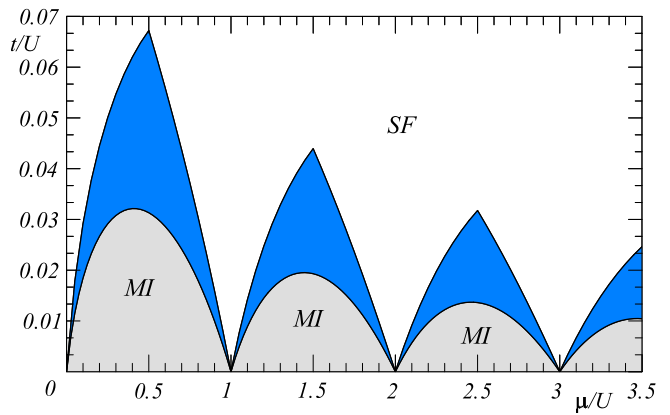}

\caption{(Color online) Phase boundary between the Mott-insulating (MI) and
superfluid (SF) phases for square (darker lobes) and cubic lattice
in the space of the parameters $t/U-\mu/U$.\label{2D3D}}
\end{figure}
The action included propagator with calculated the topological contribution
after Fourier transform we write as \begin{equation}
\mathcal{S}_{\mathrm{eff}}=\frac{1}{N\beta}\sum_{\mathbf{k},m}\psi_{\mathbf{k},m}^{*}\mathrm{\Gamma}_{\mathbf{k}}^{-1}\left(\omega_{m}\right)\psi_{\mathbf{k},m},\end{equation}
where\begin{equation}
\mathrm{\Gamma}_{\mathbf{k}}^{-1}\left(\omega_{m}\right)=\lambda-J\left(\mathbf{k}\right)+\gamma^{-1}\left(\omega_{m}\right)\label{gamma}\end{equation}
is the inverse of the propagator. Within the phase coherent superfluid
state the order parameter is given by\begin{equation}
1-\Psi_{B}^{2}=\frac{1}{N\beta}\sum_{\mathbf{k},m}\frac{1}{\lambda-J\left(\mathbf{k}\right)+\gamma^{-1}\left(\omega_{m}\right)},\label{critical line}\end{equation}
where for bipartite lattices we have:\begin{eqnarray}
J\left(\mathbf{k}\right) & = & b_{0}^{2}t\left(\mathbf{k}\right)=\left(2z\frac{t}{U}+\frac{\bar{\mu}}{U}\right)t\left(\mathbf{k}\right)\end{eqnarray}
with the dispersion $t\left(\mathbf{k}\right)$ given by Eq. (\ref{dispersion})
and $z$ is the lattice coordination number. The phase boundary is
determined by Eq. (\ref{critical line}) from the upper limit of the
eigenvalue spectrum $\mathrm{max}\left[t\left(\mathbf{k}\right)\right]$
associated with the onset of phase transition. The Lagrange multiplier
$\lambda$ ''sticks'' at criticality to the value $\lambda_{0}$
and stays constant in the whole low temperature ordered phase. The
emergence of the critical point is signaled by the condition\begin{equation}
\lambda_{0}-J\left(\mathbf{k}=0\right)+\gamma^{-1}\left(\omega_{m=0}\right)=0\end{equation}
\begin{figure}
\includegraphics[%
  scale=0.45]{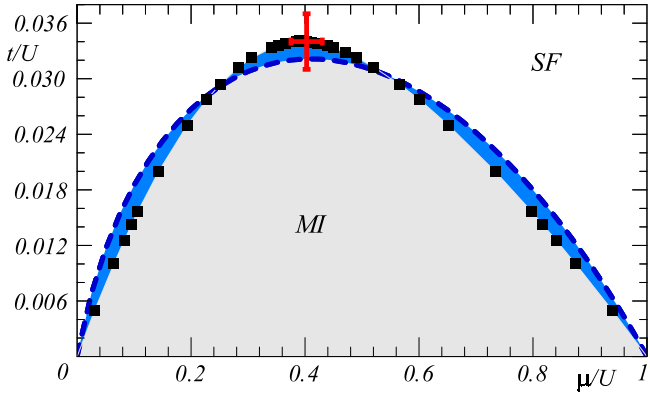}

\caption{(Color online) The comparison between our results (dashed line) and
quantum Monte Carlo (black boxes). Difference between critical values
$\left(t/U\right)_{\mathrm{crit}}$ from our theory and QMC is within
the range of error bars of the numerical calculations. The darker
area shows the difference between the critical line obtained from
Eq. (\ref{critical line final}) and phase boundary from QMC. Grey
and white areas mean the Mott-insulator and superfluid state respectively.
\label{QMC}}
\end{figure}
and by that very fact holds a converge in the constraint Eq. (\ref{critical line}).
After summation over Matsubara frequency the superfluid state order
parameter in the limit $\beta\rightarrow\infty$ becomes\begin{equation}
1-\Psi_{B}^{2}=\frac{1}{2N}\sum_{\mathbf{k}}\frac{1}{\sqrt{\frac{J\left(\mathbf{k}=0\right)-J\left(\mathbf{k}\right)}{U}+\upsilon^{2}\left(\frac{\mu}{U}\right)}}\end{equation}
with\begin{equation}
\upsilon\left(\frac{\mu}{U}\right)=\mathrm{frac}\left(\frac{\mu}{U}\right)-\frac{1}{2},\label{v parameter}\end{equation}
 where $\mathrm{frac}\left(x\right)=x-\left[x\right]$ is the fractional
part of the number and $\left[x\right]$ is the floor function which
gives the greatest integer less then or equal to $x$. Introducing
the density of states $\rho\left(\xi\right)=N^{-1}\sum_{\mathbf{k}}\delta\left[\xi-t\left(\mathbf{k}\right)\right]$
we obtain the critical line equation:\begin{equation}
1-\Psi_{B}^{2}=\frac{1}{2}\int_{-\infty}^{+\infty}\frac{\rho\left(\xi\right)d\xi}{\sqrt{2\bar{\xi}\left(2z\frac{t}{U}+\frac{\mu}{U}+\frac{1}{2}\right)\frac{t}{U}+\upsilon^{2}\left(\frac{\mu}{U}\right)}},\label{critical line final}\end{equation}
where $\xi$ is dimensionless parameter, $\bar{\xi}\equiv\xi_{\mathrm{max}}-\xi$
and $\xi_{\mathrm{max}}$ stands for the maximum value of the dispersion
spectrum $t\left(\mathbf{k}\right)$. The zero temperature phase diagram
of the model calculated from Eq. (\ref{critical line final}) is given
in Fig. \ref{2D3D}. We recognize the particle-hole asymmetric - as
a result of the model Hamiltonian Eq. (\ref{hamiltonian1}) - Mott-insulating
 lobes similar to what was found in the literature.\cite{kampf,freericks,elstner}
In the MI phase bosons are incompressible $\partial n_{B}/\partial\mu=0$
and localized which means that the total energy is minimized when
each site is filled with the same number of atoms. Increasing fluctuations
in the phase system reduces fluctuations in the boson number on each
site according to Heisenberg uncertainty relation $\Delta n_{B}\Delta\phi\geq1/2$.
Crossing the boundary line bosons can move from one lattice site to
the next. The order parameter $\Psi_{B}$ has a non-vanishing value
and system exhibits the long-range phase coherence. This is opposite
case to the Mott-insulator where phase coherence is lost. 

We found that our results are in great accordance with the recently
published quantum Monte-Carlo calculations\cite{caproso} (see Fig.
\ref{QMC}) and also improve predictions based on the third-order
expansion in $t/U$ that become inaccurate quite far from the tip.\cite{freericks} Furthermore, comparison with the strong coupling method \cite{sengupta} indicates that it underestimates the critical values of $t/U$. For example, in three dimension it gives $t/U=0.029$ at the tip of the $n=1$ lobe, which is slightly lower that the value that results form our calculation and the Monte Carlo method.
The phase boundary is periodic with respect to $\mu/U$ with fixed
integer filling depending on the value of the chemical potential $\mu$.
The vicinity of the lobe tip $\left(t/U\right)_{\mathrm{crit}}$,
corresponding to the MI-SF transition in the commensurate system is
shifting from value $0.4$ to $0.5$ when we change a dimension of
the lattice from three- $\left(3D\right)$ to two-dimensional $\left(2D\right)$.
Moreover, we see that the qualitative shape of the lobes is not the
same for $2D$ and $3D$ cases and steeper for the two-dimensional
system. Analysis of the one-dimensional systems is not possible in
presented approach because for dimensions $d\leq2$ it does not exhibit
the phase transition at finite temperatures $T>0$, in agreement with
the Mermin-Wagner theorem.\cite{mermin} 

Finally a comment regarding the critical behavior of the model in
our quantum rotor approach is in order. To extract the near-critical
form of the propagator it is this sufficient to perform an expansion
in terms of the momentum $\mathbf{k}$ and frequency $\omega_{m}$
in Eq. (\ref{gamma}). In the $T\rightarrow0$ limit, with the help
of Eq. (\ref{correlator}), after proper re-scaling of the fields
$\psi_{\mathbf{k},\omega_{m}}$ we find\begin{equation}
\mathrm{\Gamma}_{\mathbf{k}}^{-1}\left(\omega_{m}\right)=r+\mathbf{k}^{2}+\omega_{m}+i\omega_{m}+\upsilon\left(\frac{\mu}{U}\right)\label{scaling}\end{equation}
Here, $r\sim J\left(\mathbf{k}=0\right)-\lambda$ is the critical
``mass\char`\"{} parameter that vanishes at the phase transition boundary,
$\mathbf{k}^{2}=k\cdot k$ while $\upsilon\left(\frac{\mu}{U}\right)$
is given by Eq. (\ref{v parameter}). Due to the quantum nature of
the problem, the scaling of the spatial degrees of freedom $\mathbf{k}\rightarrow\mathbf{k}'=s\mathbf{k}$
implies the scaling for frequencies in a form $\omega_{m}\rightarrow\omega'_{m}=s^{z}\omega_{m}$
with the dynamical critical exponent $z$. At the tips of the lobes
in the $t/U$-$\mu/U$ phase diagram in Fig. \ref{QMC} one has $\upsilon\left(\mu/U\right)=0$,
so that $\Gamma_{{\bf k}}^{-1}(\omega_{m})\sim k^{2}+\omega_{m}^{2}$
with space-time isotropy giving $z=1$. However, the other points
on the critical line with non--vanishing $\upsilon\left(\mu/U\right)$
reflect the absence of the particle--hole symmetry due to the imaginary
term involving $i\omega_{m}$. In this case the higher order term
$\omega_{m}^{2}|\psi_{{\bf k},\omega_{m}}|^{2}$ becomes irrelevant
and can be ignored, while the critical form of the propagator (\ref{scaling})
reads $\Gamma_{{\bf k}}^{-1}(\omega_{m})\sim k^{2}+i\upsilon\left(\mu/U\right)\omega_{m}$.
Now, the scaling requires $z=2$ as a result of the momentum--frequency
anisotropy.

\subsection{Boson occupation number}

\begin{figure}
\includegraphics[%
  scale=0.4]{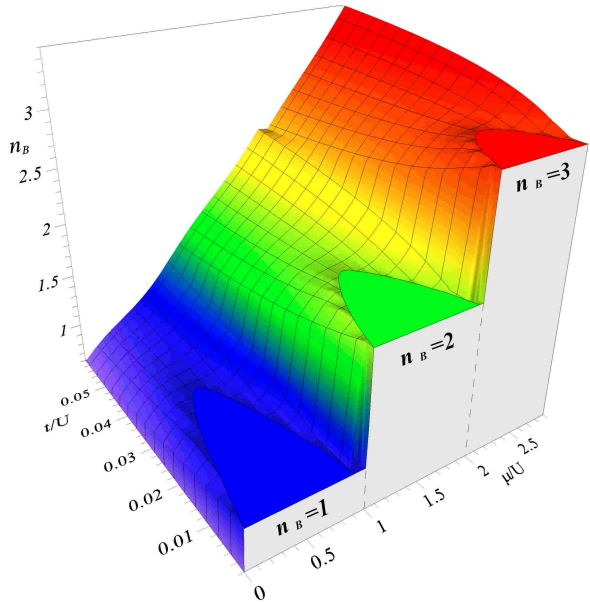}

\caption{(Color online) Boson occupation number $n_{B}$ at $T=0$ for three-dimensional
simple cubic lattice in the space of parameters - chemical potential
$\mu/U$ and hopping $t/U.$ The Mott insulator is found within each
lobe of integer boson density. Inside the first lob on the left the
occupation number $n_{B}$ is equal one, two and three in second and
third step respectively.\label{ntu}}
\end{figure}
The effects of the fixed boson number $n_{B}$ in the system defined
by \begin{equation}
n_{B}=\frac{1}{N}\sum_{i}\left\langle \bar{a}_{i}\left(\tau\right)a_{i}\left(\tau\right)\right\rangle \end{equation}
are included in our theory because of the source term containing chemical
potential $\bar{\mu}\sum_{i}\int_{0}^{\beta}d\tau\bar{b}_{i}\left(\tau\right)b_{i}\left(\tau\right)$
in action Eq. (\ref{chemical potential action}). By differentiating
the partition function Eq. (\ref{partition function}) (after carrying
out a gauge transformation and change the variable in the action)
we obtain\begin{equation}
\left\langle \bar{a}_{i}\left(\tau\right)a_{i}\left(\tau\right)\right\rangle =\frac{\partial\ln\mathcal{Z}}{\partial\bar{\mu}}=\frac{1}{iU}\left[\left\langle V_{i}^{S}\right\rangle +\left\langle \dot{\phi_{i}}\right\rangle -i\bar{\mu}\right].\end{equation}
Inserting in above a static part of the electrochemical potential
Eq. (\ref{static part}) we find the boson density\begin{equation}
n_{B}=\frac{1}{N}\sum_{i}\left[\left\langle \bar{b}_{i}\left(\tau\right)b_{i}\left(\tau\right)\right\rangle +\frac{1}{iU}\left\langle \dot{\phi_{i}}\right\rangle \right].\end{equation}
When the phase stiffness vanishes $J=0$ the bosonic contribution
to the free energy is given by\begin{equation}
\mathcal{F}\left(\bar{\mu}\right)=-\frac{1}{\beta N}\ln\int\left[\mathcal{D}\phi_{i}\right]e^{-\int_{0}^{\beta}d\tau\sum_{i}\left[\frac{1}{2U}\dot{\phi_{i}^{2}}\left(\tau\right)+\frac{1}{i}\frac{\bar{\mu}}{U}\dot{\phi_{i}}\left(\tau\right)\right]}\label{contribution bosonic}\end{equation}
which is simply the contribution from the ''free'' rotor action.
Now, we again decompose the phase field in terms of a periodic field
and term linear in $\tau$. Calculating integral Eq. (\ref{contribution bosonic})
we get in the limit $T\rightarrow0$ an analytical solution\begin{equation}
\left.n_{B}\left(\mu\right)\right|_{J=0}=\left.\frac{\partial\mathcal{F}\left(\bar{\mu}\right)}{\partial\bar{\mu}}\right|_{J=0}=\frac{\mu}{U}+\frac{1}{2}-\upsilon\left(\frac{\mu}{U}\right)\end{equation}
from which we recognize a steps of fixed integer filling of bosons
(set $t/U=0$ in the Fig. \ref{ntu}). 

The calculations of a phase diagram for interaction problem $t/U\neq0$
are more complicated since spatial correlations have to be included,
as well. However our model is expressed in terms of the complex field
Eq. (\ref{popov}) which is now very helpful. The result for the boson
density $n_{B}$ within the region of superfuidity is given by the
expression\begin{eqnarray}
n_{B} & = & \frac{\mu}{U}+\frac{1}{2}-2\Psi_{B}^{2}\upsilon\left(\frac{\mu}{U}\right),\end{eqnarray}
where non-vanishing value of the order parameter $\Psi_{B}$ is calculated
from Eq. (\ref{critical line final}). We see in the Fig. \ref{ntu}
that the competition between kinetic and interaction energy is the
foundations of the quantum phase transitions in the BH model. Increasing
the value of the hopping term (reducing the interaction energy) leads
to delocalization of the bosons thus the sharp steps of the MI state
become indistinct and in consequence system is superfluid. In Fig.
\ref{ntu} we observe the appearance of the Mott-insulating lobes
corresponding to curves from Fig. \ref{2D3D}. The MI has a gap to
density excitations and is an incompressible (density plateaus in
Fig. \ref{ntu}) thus the chemical potential can be changed within
a gap without changing the density. At the tip of the lobe at fixed
integer density the transition is driven by the change of the $t/U$
ratio in a system composed of a fixed number of bosons. Such a transition
in a $d$-dimensional BH model lies in the universality class of the
$\left(d+1\right)$-dimensional $XY$ spin model. Remain possibilities
that the system can cross the superfluid - Mott insulator phase boundary
are called generic (when we add/subtract a small number of particles)
and do not belong with the universality class of the $XY$ spin model,
so are characterized by different critical exponents.

The possibility to describe both the Mott and SF phases in two dimensions
can be also done using the strong-coupling expansion method.\cite{sengupta}
The obtained results are qualitatively comparable to our phase diagrams
and show that both extensions of the Bose-Hubbard model beyond mean-field
succeed in catching strongly interacting systems. Besides, authors
calculated the excitation energies and spectral weight and provided
analytical formulas which expanded can be useful to determine the
expected second order term in the momentum distribution.\cite{spielman}
Furthermore it seems that presented approaches can be in principle
applied to more complicated situations.

\section{Summary and outlook}

In this paper we have presented a study of the Mott-insulator transition
of the Bose-Hubbard model. To analyze a quantum phase transitions
beyond mean-field theory we employed a $\mathrm{U}\left(1\right)$
quantum rotor approach and a path-integral formulation of quantum
mechanics including a summation over a topological charge, explicitly
tailored for the BH Hamiltonian. The effective action formalism allows
us to formulate a problem in the phase only action and obtain an analytical
formulas for the critical lines. We have compared obtained results
to existing numerical calculations and found them in a very good agreement.
The formalism adopted here can be extended and applied to the other
systems systematically. Especially the effect of the competition between
quantum effects in finite temperatures focuses our attention. Considerations
different geometries of the lattices are possible in the frame of
our approach, as well. These topics will be considered in future publications.

\begin{acknowledgments}
We would like to thank Dr. Barbara Capogrosso for an access to the
numerical data and valuable discussions. One of us (T.K.K) acknowledges
the support by the Ministry of Education and Science MEN under Grant
No. 1 P03B 103 30 in the years 2006-2008.
\end{acknowledgments}

\end{document}